\documentclass[twocolumn, times]{aastex61}

\accepted{\today}
\submitjournal{ApJ Supplement}

\shorttitle{Hydrogen Dimers on Giant Planets}
\shortauthors{Fletcher et al.}

\begin{document}

\title{Hydrogen dimers in giant-planet infrared spectra}

\correspondingauthor{Leigh Fletcher}
\email{leigh.fletcher@le.ac.uk}

\author{Leigh N. Fletcher}
\affil{Department of Physics and Astronomy, University of Leicester, University Road, Leicester, LE1 7RH, UK.}

\author{Magnus Gustafsson}
\affiliation{Applied Physics, Division of Materials Science, Department of Engineering Science and Mathematics, Lule{\aa} University of Technology, SE-97187 Lule{\aa}, Sweden}

\author{Glenn S. Orton}
\affiliation{Jet Propulsion Laboratory, California Institute of Technology, 4800 Oak Grove Drive, Pasadena, CA 91109, USA}



\begin{abstract}


Despite being one of the weakest dimers in nature, low-spectral-resolution Voyager/IRIS observations revealed the presence of (H$_2$)$_2$ dimers on Jupiter and Saturn in the 1980s.  However, the collision-induced H$_2$-H$_2$ opacity databases widely used in planetary science \citep{85borysow, 07orton, 12richard} have thus far only included free-to-free transitions and have neglected the contributions of dimers.  Dimer spectra have both fine-scale structure near the S$(0)$ and S$(1)$ quadrupole lines (354 and 587 cm$^{-1}$, respectively), and broad continuum absorption contributions up to $\pm50$ cm$^{-1}$ from the line centres.  We develop a new \textit{ab initio} model for the free-to-bound, bound-to-free and bound-to-bound transitions of the hydrogen dimer for a range of temperatures (40-400 K) and para-hydrogen fractions (0.25-1.0).  The model is validated against low-temperature laboratory experiments, and used to simulate the spectra of the giant planets.  The new collision-induced opacity database permits high-resolution (0.5-1.0 cm$^{-1}$) spectral modelling of dimer spectra near S$(0)$ and S$(1)$ in both Cassini Composite Infrared Spectrometer (CIRS) observations of Jupiter and Saturn, and in Spitzer Infrared Spectrometer (IRS) observations of Uranus and Neptune for the first time.  Furthermore, the model reproduces the dimer signatures observed in Voyager/IRIS data near S$(0)$ \citep{84mckellar} on Jupiter and Saturn, and generally lowers the amount of para-H$_2$ (and the extent of disequilibrium) required to reproduce IRIS observations.


\end{abstract}




\section{Introduction} \label{sec:intro}

Far-infrared spectra of giant planet atmospheres are dominated by continuum absorption resulting from collisions between molecular hydrogen and helium.  This H$_2$-H$_2$ and H$_2$-He collision-induced absorption (CIA) provides a thermometer to measure the thermal structure of the upper tropospheres of Jupiter, Saturn, Uranus and Neptune \citep[e.g.][]{98conrath}, and has been exploited by Earth-based astronomers, space observatories (e.g., ISO, Spitzer, AKARI, Herschel), and visiting spacecraft (e.g., Voyager, Galileo, and Cassini).  The short-lived dipoles induced by these collisions between free molecules (\textit{free-to-free} interactions) generate broad and smooth spectral features around the rotational S$(0)$ (354 cm$^{-1}$) and S$(1)$ (587 cm$^{-1}$) lines.  Transitions between para-H$_2$ rotational levels $0\rightarrow2$ are responsible for the S$(0)$ features (even rotational quantum numbers), and transitions between ortho-H$_2$ rotational levels $1\rightarrow3$ are responsible for the S$(1)$ feature.  Measurement of the far-infrared spectrum can therefore provide constraints on the tropospheric temperatures, para-H$_2$ fractions \citep[a tracer of atmospheric mixing,][]{98conrath}, and the bulk helium abundance \citep{00conrath}. 

It has been common practice to only include these free-to-free contributions when calculating the opacity of giant-planet atmospheres, in addition to the narrow quadrupole lines.  However, sharp features near S$(0)$ (354 cm$^{-1}$) identified in Voyager IRIS 4.3-cm$^{-1}$-resolution spectra of Jupiter and Saturn were attributed to hydrogen dimers by \citet{84frommhold} and \citet{84mckellar}, and confirmed via low-temperature experiments by \citet{88mckellar}.  This dimeric absorption, detectable as a slight modification to the free-free absorption of the parent molecules, is the result of the formation of a weakly-bound (H$_2$)$_2$ complex held together by van-der-Waals forces.  The main mechanism of dimer formation and dissociation is three-body interactions, which sustain a dimer abundance determined by thermal equilibrium. To a much smaller extent, dimers may be formed during \textit{free-to-bound} (H$_2$+H$_2\rightarrow$ (H$_2$)$_2$) radiative transitions, which generate features redward (smaller wavenumbers) of the rotational line centre in the absorption spectrum; and dissociated during \textit{bound-to-free} transitions ((H$_2$)$_2$ $\rightarrow$ H$_2$+H$_2$) that generate spectral signatures blueward (larger wavenumbers) of the line centre \citep[e.g., from ab initio models by][]{84frommhold}.  As the dimer lifetime is much longer than the short-lived free-to-free collisions, the spectral features are much narrower.  Despite being one of the weakest dimers in nature, the (H$_2$)$_2$ contribution was readily visible in the Voyager spectra near S$(0)$.  

Following their identification, a series of experimental and theoretical results were presented to characterise the dimer contributions \citep{87schaefer, 88mckellar, 90mckellar, 91mckellar}.  The strength of the dimer absorption depends on the abundance of H$_2$, the ortho-para ratio, and the temperature.  The influence of the dimers on the spectrum is most important at low temperature, as shown in the 20-K experiments by \citep{91mckellar}.  \citet{89meyer} demonstrated that the extra free-to-bound and bound-to-free transitions could alter the continuum tens of wavenumbers away from the line centre.  \citet{92carlson} fitted the Meyer calculations with empirical formulae to incorporate this broadband dimer structure into the IRIS modelling, finding that this additional absorption significantly improved their fit near 350 cm$^{-1}$.  Furthermore, \citet{95kim} and \citet{97trafton} identified dimer emission in Jupiter, Saturn and Neptune near the H$_2$ fundamental at 2.1 $\mu$m.  


However, despite these pioneering studies, subsequent analyses over the past two decades have largely omitted this additional absorption and considered only the free-to-free transitions.  Analyses of giant planet infrared spectra have typically utilised the free-free H$_2$-H$_2$ opacity model of \citet{85borysow}, which was based on \textit{ab initio} dipole surfaces from \citep{89meyer}.  However, \citet{07orton} demonstrated that an error had been made in the modelling of the dipole components by \citet{85borysow}, which appeared at the rotational double transitions in the absorption spectrum.  \citet{07orton} carried out extensive corrected calculations, which were otherwise at the same level of theory as \citet{85borysow}.   Their corrected free-to-free coefficients, which were only provided (i) for normal H$_2$ (i.e., a 3:1 mixture of ortho-H$_2$ and para-H$_2$) and (ii) for H$_2$ with para and ortho states in equilibrium at the local temperature, now form the `Alternative' CIA database maintained by HITRAN \citep{12richard}.  However, \citet{17fletcher_sofia} showed that this new free-to-free calculation resulted in smaller absorption coefficients near the S$(0)$ and S$(1)$ lines than the original database of \citet{85borysow}, which led to spurious effects when fitting Voyager/IRIS spectra of Jupiter, despite being a more accurate calculation of the free-to-free contribution.  The solution requires the addition of the dimer opacity, as originally envisaged by \citet{84frommhold}, \citet{92carlson} and others.  This was employed by \citet{14orton} in their analysis of Uranus' S$(1)$ line from Spitzer data, which confirmed the existence of dimer features in Uranus' far-infrared spectrum for the first time.  In summary, use of the free-to-free absorption coefficients of \citet{07orton} in isolation will underestimate the opacity required to properly reproduce giant planet spectra.  Bound-to-free and free-to-bound absorption coefficients, calculated for a range of temperatures and para-H$_2$ fractions, are also required.

Section \ref{calc} presents new calculations of the free-to-free, bound-to-free, free-to-bound and bound-to-bound ((H$_2$)$_2\rightarrow$ (H$_2$)$_2$) contributions to the H$_2$-H$_2$ opacity for giant planet atmospheres, extending the work of \citep{07orton}.  Section \ref{model} shows that the new \textit{ab initio} model can be used to reproduce the fine-scale structure observed in high-resolution spectroscopy of the S$(0)$ and S$(1)$ regions on all four giant planets, using data from Cassini (Jupiter and Saturn) and Spitzer (Uranus and Neptune).  We present the first observations of the S$(0)$ dimer on Uranus; the first observation of both S$(0)$ and S$(1)$ dimers on Neptune; and the first observation of the S$(1)$ dimers on Jupiter and Saturn.  Inclusion of this structured dimer absorption along with the smooth CIA will be essential for interpretations of spectra from the James Webb Space Telescope.



 
\section{Spectral Calculation}
\label{calc}

The quantum mechanical calculations of the interaction-induced spectra have been divided into three categories according to the physical mechanism, each category having its own method of calculation of the H$_2$--H$_2$ wave functions.  The first two categories are also described briefly in \citet{17gustafsson}.

\textit{Free-to-free:}  This contribution is computed using the conventional method with an isotropic interaction potential \citep{89meyer}.  The use of the isotropic potential approximation (IPA) is justified as the anisotropy has been shown to have only a small effect on the absorption coefficient for H$_2$--H$_2$, in particular around the rotational transitions that we focus on in this work \citep{03gustafsson,15karman}.  A Numerov algorithm \citep{68korn} is implemented to obtain the one-dimensional continuum wave functions.  The table presented by \citet{07orton} has been refined in the regions surrounding the S(0) and S(1) transitions, 310-400 cm$^{-1}$ and 530-630 cm$^{-1}$, respectively, to account for the presence of fine features in the energy-dependent absorption cross section.  These have to be resolved for an accurate numerical integration over energy, which is done to obtain the temperature-dependent absorption coefficient.  Finally, we extend the work of \citet{07orton} by calculating the free-to-free absorption coefficients on a grid of ten temperatures (40-400 K) and ten para-H$_2$ fractions (0.25-1.0).  

\textit{Bound-to-free and free-to-bound:} Here the isotropic potential approximation (IPA) is also applied.  The bound state wave functions are computed with a discrete variable representation (DVR) with a uniform grid \citep{92colbert} and the continuum states are computed with the Numerov algorithm. The formula for the absorption coefficient is taken from \citet{89meyer}.  A 0.5-cm$^{-1}$ empirical shift of the upper dimer level was applied for the free-to-bound transitions on the low-frequency side of the S(0) line.   The shift is applied for collisions where the H$_2$ molecules are in the $(j_1, j_2) = (0, 0), (0, 1)$, or $(1, 0)$ rotational states.  The shift was identified empirically via comparison with experimental data, but the magnitude is consistent with the comparison of these dimer levels (which are computed with the isotropic potential) with those computed using the full anisotropic potential, given below.  For the upper dimer level the difference is on the order of 0.5 cm$^{-1}$.  

\textit{Bound-to-bound:}  These calculations require the inclusion of the anisotropy of the potential to give transition frequencies in agreement with those observed in the laboratory.   A new program, using the DVR algorithm with an anisotropic potential, was developed to compute the absorption coefficient. The method is outlined in Appendix~\ref{appendix_bb}.

All calculations have been carried out using the potential by \citet{89schafer} and the dipole by \citet{89meyer_b}.  A more recent potential surface, which was used by \citet{15karman}, has also been tested, but it failed to give dimer states that are consistent with the experiments.

\begin{figure*}[ht!]\epsscale{0.9}
\plottwo{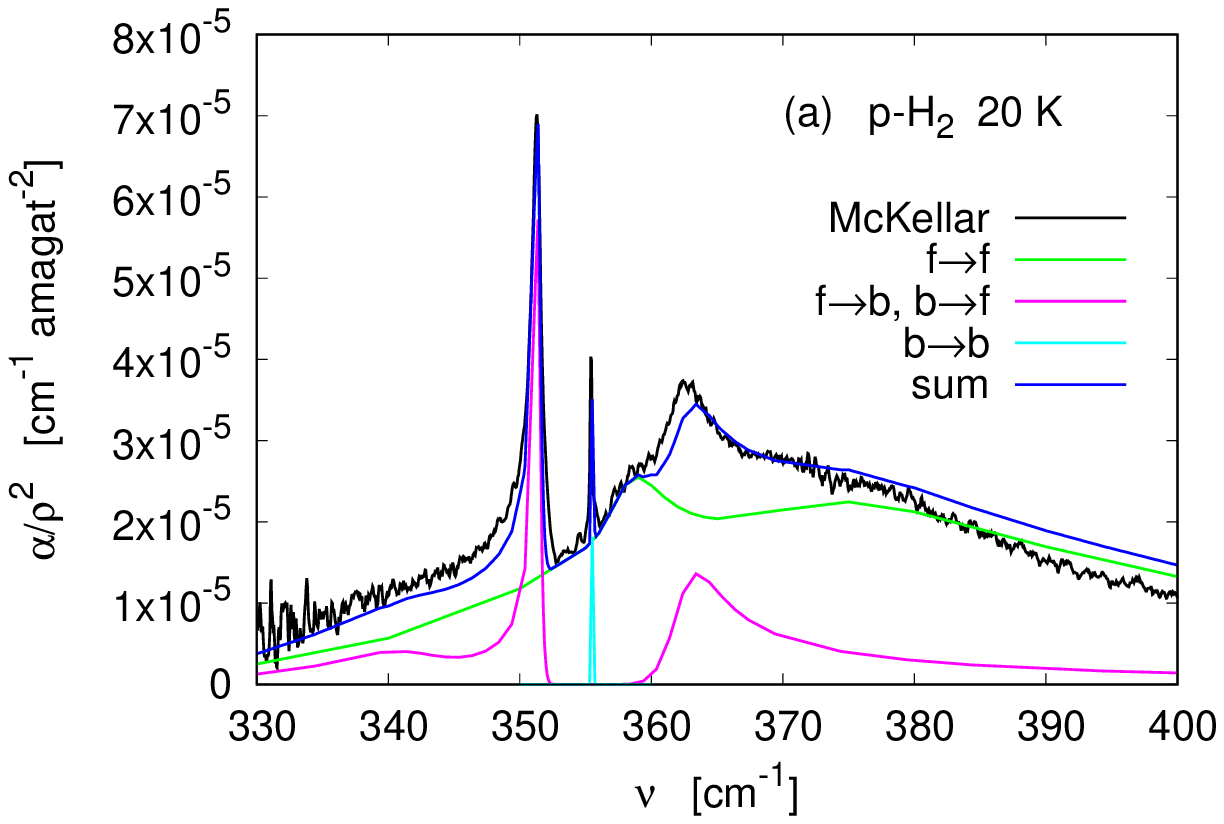}{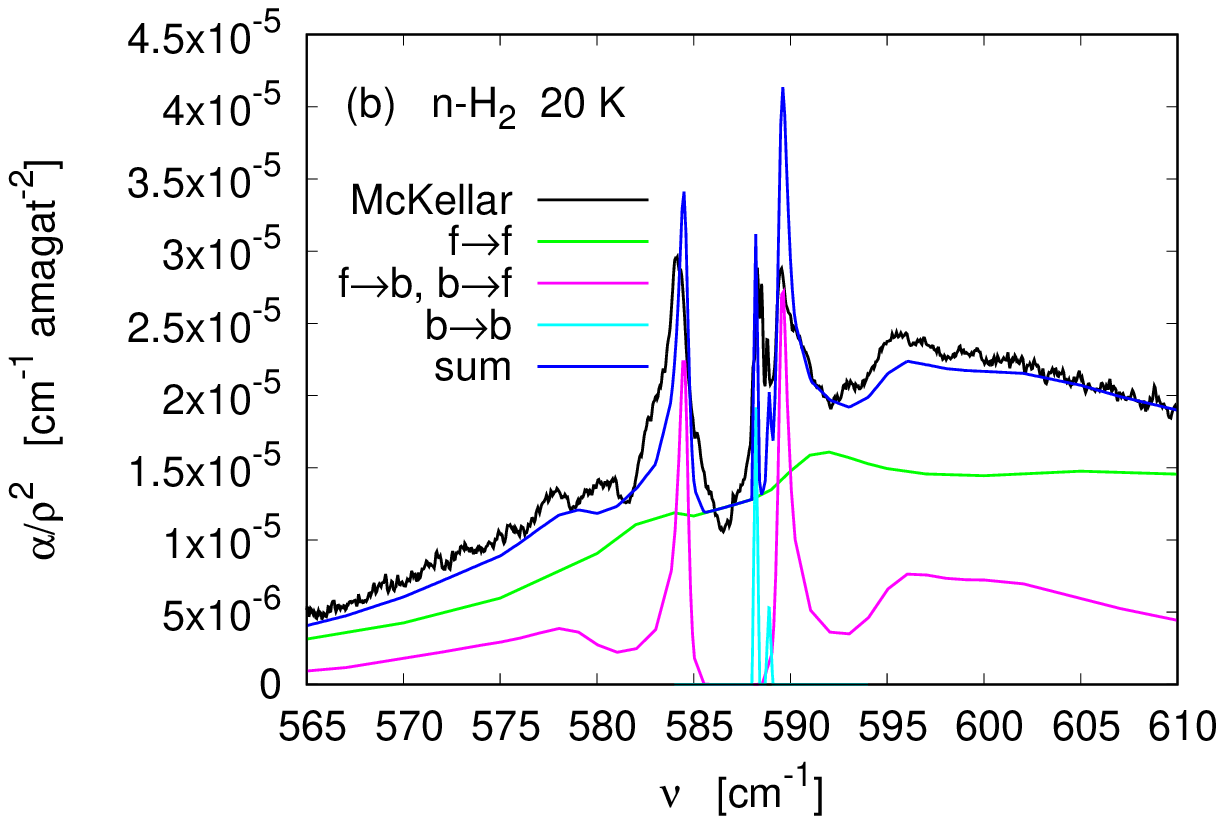}
\caption{The absorption coefficient at 20~K, normalised by the square of the hydrogen density, around (a) the S(0) and (b) the S(1) transitions for pure para-hydrogen and for normal-hydrogen, respectively.  The calculations are performed as described in Section~\ref{calc}.  The line labeled `McKellar' represents the laboratory measurements by \citet{91mckellar}.  Pressures of about 49 Torr (equivalent to a density of 0.88 amagat) and 40 Torr (equivalent to a density of 0.72 amagat) were used in experiments in (a) and (b), respectively.
\label{fig:abs20K}}
\end{figure*}

\begin{figure*}[ht!]\epsscale{0.9}
\plottwo{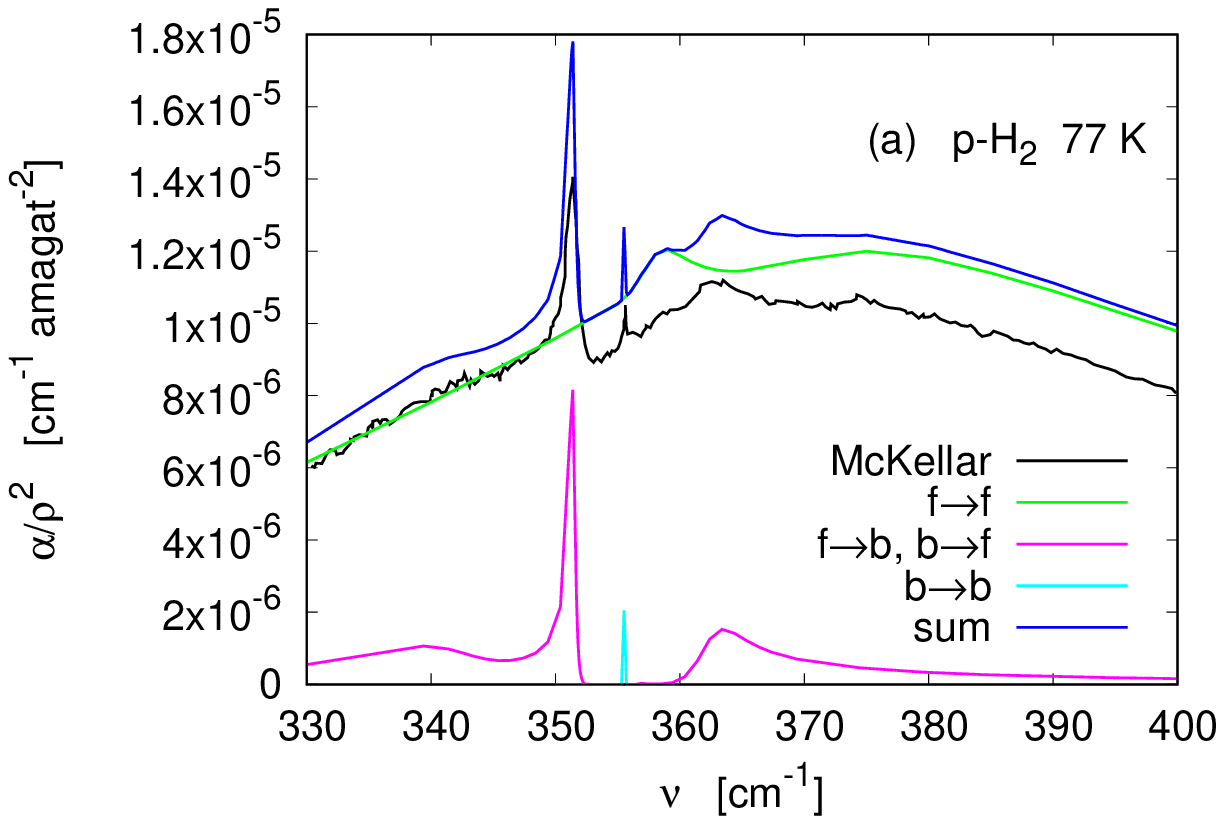}{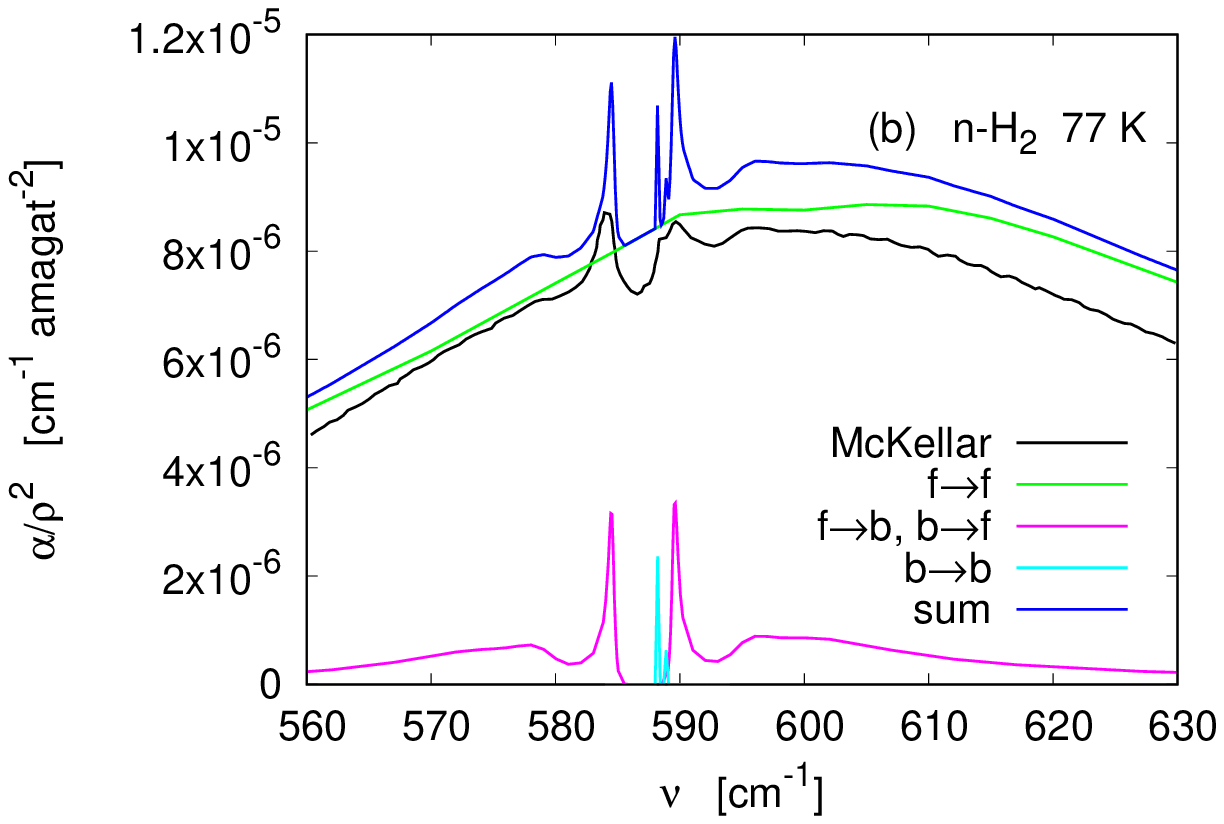}
\caption{As in Fig.~\ref{fig:abs20K} but at a temperature of 77~K.  In this case the laboratory measurements are from \citet{88mckellar} and were taken at number densities of 1.30 amagats and 2.45 amagats for (a) and (b), respectively.\label{fig:abs77Kpn}}
\end{figure*}

\begin{figure*}[ht!]\epsscale{0.9}
\plottwo{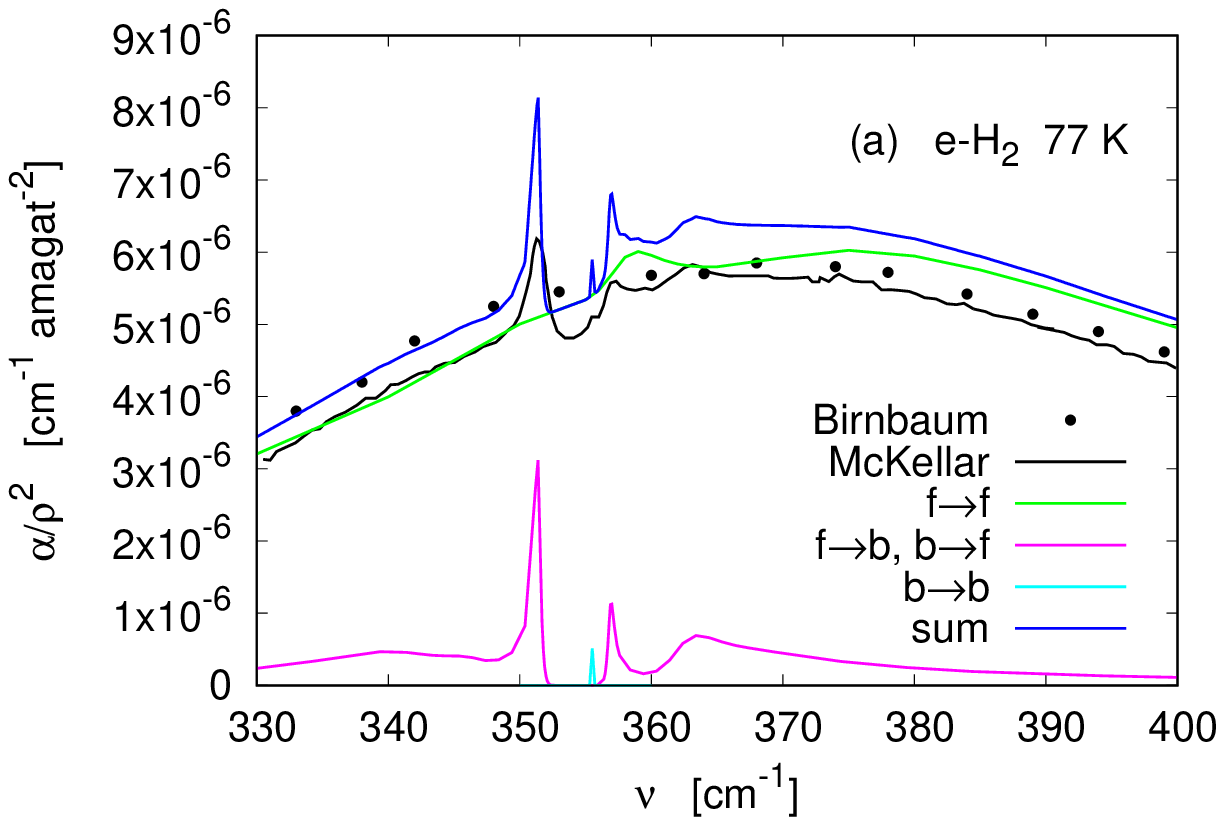}{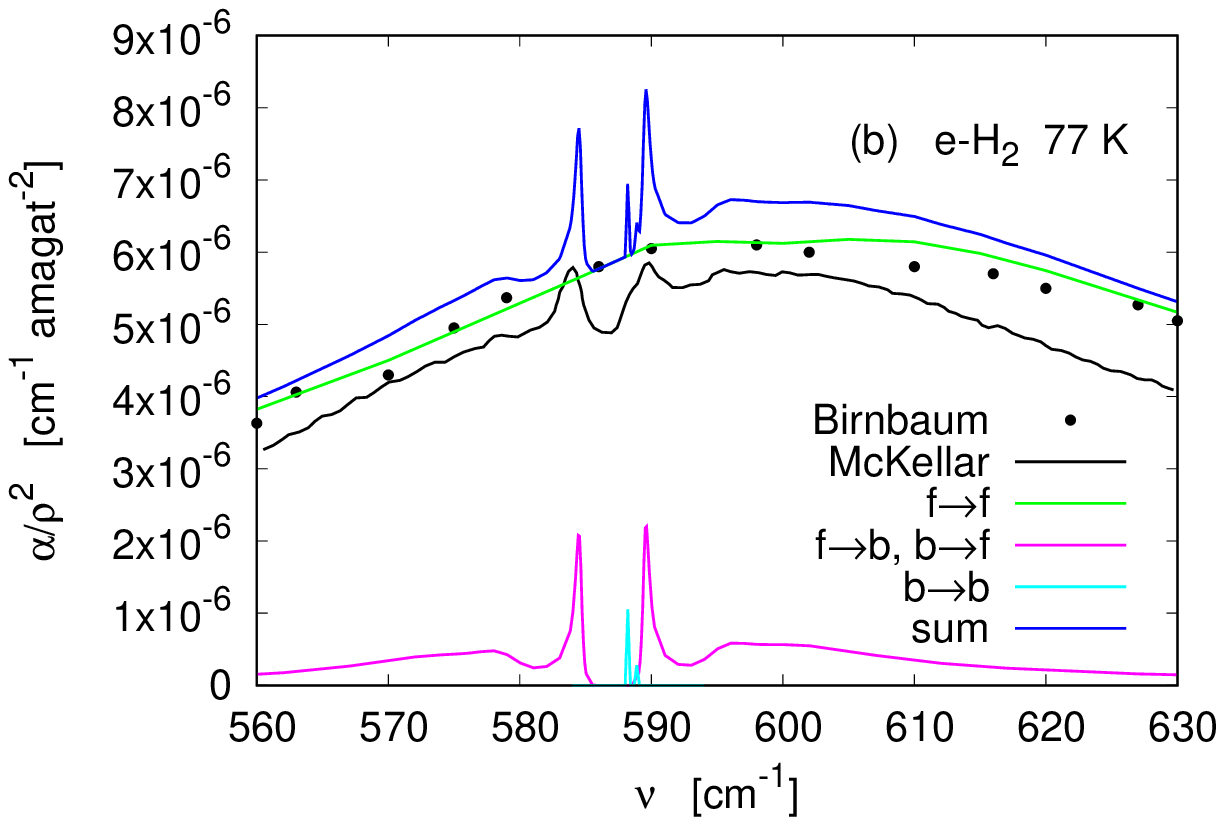}
\caption{As in Fig.~\ref{fig:abs77Kpn} but with equilibrium-hydrogen in both (a) and (b).  The laboratory measurement by \citet{88mckellar} was taken at a number density of 2.63 amagats.  Additional low-resolution laboratory measurements by \citet{78birnbaum} are also included (circles), but these do not resolve the dimer features.  However, it confirms that the overall magnitude of the calculated absorption is consistent with the experimental measurements.
\label{fig:abs77Ke}}
\end{figure*}

Figs. \ref{fig:abs20K}, \ref{fig:abs77Kpn}, and \ref{fig:abs77Ke} compares our calculated absorption spectra surrounding the S(0) and S(1) transitions with experimental data at temperatures of 20~K \citep{91mckellar} and 77~K \citep{88mckellar}.  A variety of para- to ortho-hydrogen ratios have been used in these figures.  The computed bound-to-bound spectrum was convolved with a triangular slit function of width $w$= 0.2 cm$^{-1}$ to match the experimental resolution.  The sums of the spectral contributions agree reasonably well with the experiment.  For the cases of pure para-hydrogen the agreement is particularly good (Figs. \ref{fig:abs20K}-\ref{fig:abs77Kpn}).  For the case of equilibrium-hydrogen (Fig. \ref{fig:abs77Ke}) the agreement is better for the S(0) transition than for the S(1) transition.  This difference implies that the $j_i > 0$ monomer rotations are more poorly described in the bound-to-free and free-to-bound cases, indicating that an anisotropic potential treatment would improve the agreement.   We also note that the density was higher for the experimental measurements taken with a larger fraction of ortho-hydrogen, so that pressure broadening may have affected these spectral features.

\section{Giant-Planet Modelling}
\label{model}
 
\subsection{Sources of Data}

To date, the (H$_2$)$_2$ dimer absorption has only been studied near the S$(0)$ feature in Voyager/IRIS 4.3-cm$^{-1}$ resolution spectra of Jupiter and Saturn \citep{84frommhold} and were identified in $R\sim600$ resolution Spitzer spectra of the S$(1)$ feature on Uranus \citep[][, although no attempts were made to fit the spectral features]{14orton}.   We utilise Cassini Composite Infrared Spectrometer \citep[CIRS,][]{04flasar} spectra of Jupiter and Saturn with a wavenumber-independent 0.48-cm$^{-1}$ spectral resolution, and Spitzer Infrared Spectrometer \citep[IRS,][]{04houck} wavelength-dependent $R\sim600$ spectra of Uranus and Neptune (i.e., 0.59 cm$^{-1}$ resolution at S$(0)$ and 0.98 cm$^{-1}$ resolution at S$(1)$).  However, as the true spectral resolution of the high-resolution Spitzer modes is uncertain, we find that a resolution of 0.48 cm$^{-1}$ is sufficient for our calculations. 

\begin{figure*}[ht!]
\plotone{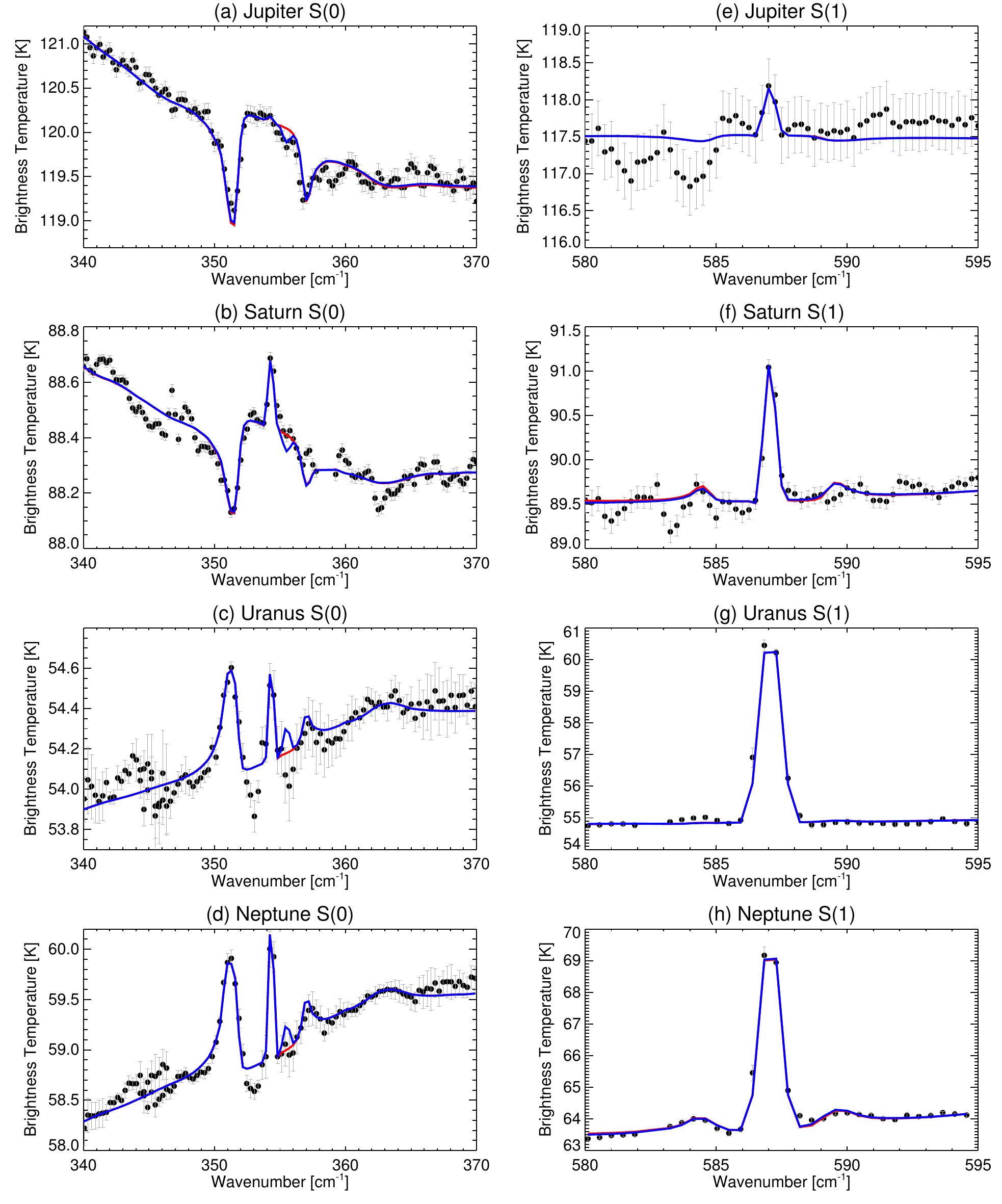}
\caption{Spectral fits to the S(0) and S(1) lines detected on all four giant planets, using Cassini data for Jupiter and Saturn and Spitzer data for Uranus and Neptune.}
\label{compareS0}
\end{figure*}

\textit{Jupiter and Saturn: }  Spectra of the S$(0)$ line were acquired by the far-IR polarising CIRS focal plane one with its circular 4.3-mrad diameter field of view, whereas spectra of the S$(1)$ line used the 0.273-mrad square detectors of CIRS focal plane 3.  Given that the detector responsivities are low in these spectral regions, large numbers of spectra were coadded to generate a single average.  We ensured that the footprints of the detectors were fully on the planetary disc, and that all spectra with emission angles smaller that 45$^\circ$ were averaged.  For Jupiter, we averaged spectra spanning from November 15th 2000 - February 15th 2001 during the Cassini flyby - 850 spectra were used for S$(0)$, 15,000 spectra for S$(1)$.  For Saturn, we averaged spectra from October 2004 to December 2016 with the same criteria, using 57,000 spectra for S$(0)$ and 86,000 spectra for S$(1)$.  Although these large averages were necessary to improve the signal-to-noise, the resulting atmospheric profiles (temperature and para-H$_2$) are averaged over a broad region of the planet and, in the case of Saturn, over different seasons from southern summer to northern spring.  As shown in Section \ref{discuss} and Fig. \ref{compareS0}, these spectra reveal dimeric structure around S$(0)$ on both planets and, tentatively, around S$(1)$ on Saturn.

\textit{Uranus and Neptune: }  Spitzer/IRS acquired 7-36 $\mu$m disc-integrated spectra of Uranus during Directors Discretionary Time on December 16-17, 2007 shortly after Uranus' equinox \citep[program 467, full details of the data reduction process are provided by][]{14orton}.  Disc-integrated spectra of Neptune were acquired on November 19-20, 2005 during Spitzer's Cycle 2 (program 20500), and reduced using the same process as \citet{14orton}.  For both planets, the observations were designed to sample multiple longitudes during a complete rotation, but we averaged all longitudes to form a single spectrum.  Here we focus on high-resolution ($R\sim600$) observations in the `Short-High' (SH) 9.95-19.30 $\mu$m and `Long High' (LH) 19.27-35.97 $\mu$m ranges, revealing the S$(0)$ and S$(1)$ lines in Fig. \ref{compareS0}, respectively.  Note that \citet{14orton} identified offsets between low- and high-resolution Uranus spectra related to flux losses from different slit sizes, and ultimately scaled the SH data to match the low-resolution modes, and abandoned the LH data from their analysis entirely.  In the present work, where only narrow spectral ranges are considered to identify the dimer features, we do not find it necessary to perform such a scaling and present good fits to both SH and LH data.

\textit{Voyager/IRIS: }  To supplement the high-resolution Cassini and Spitzer observations, we also reanalyse the low-resolution (4.3-cm$^{-1}$) IRIS observations of all four giant planets.  These 180-2500 cm$^{-1}$ spectra were acquired by a Michelson interferometer in a series of north-south scans executed during the close flybys between 1979 and 1989.  The selection criteria for the spectra and zonal averaging techniques have been previously described for Jupiter \citep{17fletcher_sofia}, Saturn \citep{16fletcher}, Uranus \citep{15orton} and Neptune \citep{14fletcher_nep}.  Dimer features were previously identified only in the Jupiter and Saturn spectra \citep{84frommhold}, with the Uranus and Neptune data being too low in signal to reveal the dimer absorptions.  The zonally-averaged 300-550 cm$^{-1}$ regions of these spectra are re-analysed with the new opacity database in Section \ref{discuss}.

\subsection{Spectral Retrieval Model}

The NEMESIS spectral retrieval algorithm \citep{08irwin} was used to model the H$_2$ quadrupoles, dimers and continuum absorption on all four giant planets in Fig. \ref{compareS0}.  NEMESIS uses Newtonian iteration and an optimal estimation retrieval architecture \citep{00rodgers} to calculate planetary spectra, maximising the quality of the spectral fit whilst using constraints to prior data (in this case, vertical temperature and para-H$_2$ profiles) to ensure smooth and physically-realistic retrieved profiles.  For the spatially-resolved spectra of Jupiter and Saturn, NEMESIS was used in its standard mode, computing spectra using the mean geometry (latitude and emission angle) of the spectral averages \citep[e.g.,][]{09fletcher_ph3}.  For the disc-integrated Uranus and Neptune spectra, we employed an exponential-integral technique \citep{89goody} to compute the radiance into a hemisphere \citep{14fletcher_nep}. Although the spectra explored in this paper are formed primarily by H$_2$, He (and H$_2$-CH$_4$ collisions, to a lesser extent), we use full atmospheric priors for each planet with tropospheric (CH$_4$, PH$_3$, NH$_3$) and stratospheric species (hydrocarbons) included based on previous NEMESIS studies of Jupiter and Saturn \citep{09fletcher_ph3}, Uranus \citep{14orton, 14orton_chem} and Neptune \citep{14fletcher_nep}.   Although not strictly necessary in this analysis, the sources of spectral linedata for these species are located in Table 4 of \citet{12fletcher}.

The new \textit{ab initio} model described in Section \ref{calc} produced calculations of the interaction-induced dimer spectra (free-to-bound, bound-to-free, bound-to-bound).  This was combined with updated estimates of the free-to-free H$_2$-H$_2$ contribution, the H$_2$-He contribution from \citet{88borysow}, and H$_2$-CH$_4$ and CH$_4$-CH$_4$ contributions from \citet{86borysow,87borysow}.  Extending the work of \citet{07orton}, the new free-to-free H$_2$ and He contributions, as well as the dimer contributions, were calculated for a range of para-H$_2$ fractions between 0.25 and 1.0, and for temperatures from 40 to 400 K.  These tables are then interpolated during the temperature and para-H$_2$ retrievals to calculate the atmospheric transmission.  In addition, we include the H$_2$ quadrupole transitions from the \textit{ab initio} calculations described in \citet{13rothman}, which are in agreement with recent experimental results of \citet{12campargue}.  Specifically, S$(0)$ occurs at 354.3732 cm$^{-1}$ with an intensity of $1.664\times10^{-28}$ cm$^{-1}$/molec cm$^{-2}$; S$(1)$ occurs at 587.0320 cm$^{-1}$ with an intensity of $2.657\times10^{-27}$ cm$^{-1}$/molec cm$^{-2}$.  Both have a width of 0.0017 cm$^{-1}$/atm \citep{94reuter} and temperature dependence $T^n$ where $n=0.75$.  Full vertical profiles of $T(p)$ and para-H$_2$ ($f_p(p)$) were retrieved independently to reproduce the eight spectral regions shown in Fig. \ref{compareS0}.  Note that retrievals over a broader spectral range would be required to fully constrain these atmospheric profiles, whereas the present study aims to show that the fine-scale dimeric structure can be adequately reproduced.


\section{Results and Discussion}
\label{discuss}

\subsection{High-resolution dimer structure}

Fig. \ref{compareS0} compares model spectra to the high-resolution Cassini and Spitzer measurements in the regions within a few tens of wavenumbers of the quadrupole transitions.  On Jupiter and Saturn, the CIRS data provided better signal-to-noise near the S$(0)$ line than the S$(1)$ line, permitting identification of the main free-to-bound (351.1 cm$^{-1}$, $l=2\rightarrow1$) and bound-to-free (357 cm$^{-1}$, $l=1\rightarrow2$) transitions $\sim3$ cm$^{-1}$ either side of the quadrupole.  These features are seen in absorption, as the strongest dimer absorptions are sensing higher, cooler altitudes near to the tropopause on both planets.  The asymmetry in the line intensities between these two features (shown in Fig. \ref{fig:abs20K}-\ref{fig:abs77Ke} for the S$(0)$ line) is related to the boson symmetry of para-H$_2$ \citep{84frommhold}.  Additional undulations in the continuum can be seen near 346 cm$^{-1}$ and 363 cm$^{-1}$ (free-to-bound $l=3\rightarrow0$ and bound-to-free $l=0\rightarrow3$, respectively) that contribute to the overall absorption provided by the broad free-free transition.  Although these gas-giant S$(0)$ dimers have been previously studied at low resolution from Voyager/IRIS \citep[see Section \ref{iris},][]{84frommhold, 84mckellar}, this is the first observation at a sufficient spectral resolution to resolve their line shapes.

The CIRS observations near 587 cm$^{-1}$ have much poorer noise characteristics near the edge of the detector responsivity curve, so little is identifiable beyond the central quadrupole lines.  Fig. \ref{compareS0}e-f shows that the expected dimer structure is within the uncertainty on the measurement, but we expect jovian absorption and saturnian emission near 584.5 cm$^{-1}$ and 589.5 cm$^{-1}$, $\sim2.5$ cm$^{-1}$ either side of the quadrupole emission.  Saturn's S$(1)$ dimer appears in emission rather than absorption because it senses altitudes just above the tropopause, in the region where temperature begins to increase with altitude in the lower stratosphere.

The general appearance of the dimer absorption near S$(0)$ changes character considerably for Uranus and Neptune, where both the data and model indicate that the main free-to-bound and bound-to-free transitions $\sim3$ cm$^{-1}$ from the line centre are sensing stratospheric altitudes, warmer than the surrounding free-free transitions.  The 351-cm$^{-1}$ feature almost matches the intensity of the quadrupole line itself.   The measured spectrum becomes increasingly noisy shortward of 347 cm$^{-1}$ due to an overlap of modes in the Spitzer LH setting.  However, this cannot account for the extremely poor fit to the spectrum between 352-354 cm$^{-1}$, where no dimer transitions exist to add to the absorption of the free-free contribution.  The data suggest that excess absorption could be required to make the spectrum sense higher, cooler altitudes towards the tropopause, but experiments with \textit{ad hoc} modifications to the dimer database failed to offer improvements.  Given that there is no evidence of excess absorption in the gas giant spectra, and that there are additional deviations in the Spitzer ice giant measurements that aren't accounted for by our model, it is likely that this mismatch is unrelated to the dimer spectra and is simply the result of noisy data.  New measurements of the dimer features with JWST (see Section \ref{conclusion}) or other far-IR facilities such as SOFIA should help to resolve this conundrum.

Ice giant dimer features near S$(1)$ are significantly weaker, owing to the quasi-isothermal tropopause regions to which the dimers are sensitive.  Uranus' S$(1)$ dimers were first identified by \citet{14orton}, but no attempt was made to fit the discrete structures.  In Fig. \ref{compareS0}g, the free-to-bound transition near 584.5 cm$^{-1}$ is present in the data, but our model struggles to fit it due to the weakness of the bound-to-free transition near 589.5 cm$^{-1}$ (the spectral fit is a compromise over fitting the whole region). The signal-to-noise ratio of the Spitzer data is excellent here, so this could represent a deficiency of the model.  However, the model is more successful at fitting Neptune's S$(1)$ line and dimer structure, which is stronger than for Uranus and shows equal strength in the free-to-bound and bound-to-free transitions.  We conclude that the new \textit{ab initio} model provides adequate spectral fits to S$(0)$ and S$(1)$ features on all four giant planets, despite some model-data discrepancies that we hope to constrain with improved future measurements.

Finally, we note that two spectral models are shown in Fig. \ref{compareS0} - one without bound-to-bound transitions (red) and one with them (blue). Although the bound-to-bound transitions were clearly present in the experimental data (Fig. \ref{fig:abs20K}-\ref{fig:abs77Ke}), their influence near S$(1)$ is negligible, and their importance on the giant planet S$(0)$ spectra is unclear.  On Jupiter, the addition of the bound-bound opacity does improve the fit near 355.5 cm$^{-1}$.  However, on Saturn, Uranus, and Neptune, the inclusion of the bound-bound data actually worsens the fits.  We have chosen to show both options, although the requirement for the bound-bound contribution remains unclear.

\subsection{Low-resolution:  Voyager/IRIS}
\label{iris}

Spectral modelling of Voyager/IRIS 300-550 cm$^{-1}$ spectra of Jupiter by \citet{17fletcher_sofia} had revealed a problem when the free-to-free transitions of \citet{07orton} were used in isolation, as they under-predicted the amount of collision-induced opacity required to fit the data with a physically-plausible atmospheric structure.  This present study was prompted by the need to add dimer opacity to the free-free continuum to explain the Voyager data.  Ironically, the error in the calculation of some of the free-free components by \citet{85borysow} had led to an opacity dataset that was closer to reality.  We repeat the Voyager/IRIS fitting using three assumptions for the H$_2$-H$_2$ opacity: (i) the original free-free database of \citet{85borysow}, (ii) the newer free-free database of \citet{07orton}, and (iii) the complete free-to-free, free-to-bound, bound-to-free and bound-to-bound database of this work.  

\begin{figure*}[ht!]
\includegraphics[width=\linewidth]{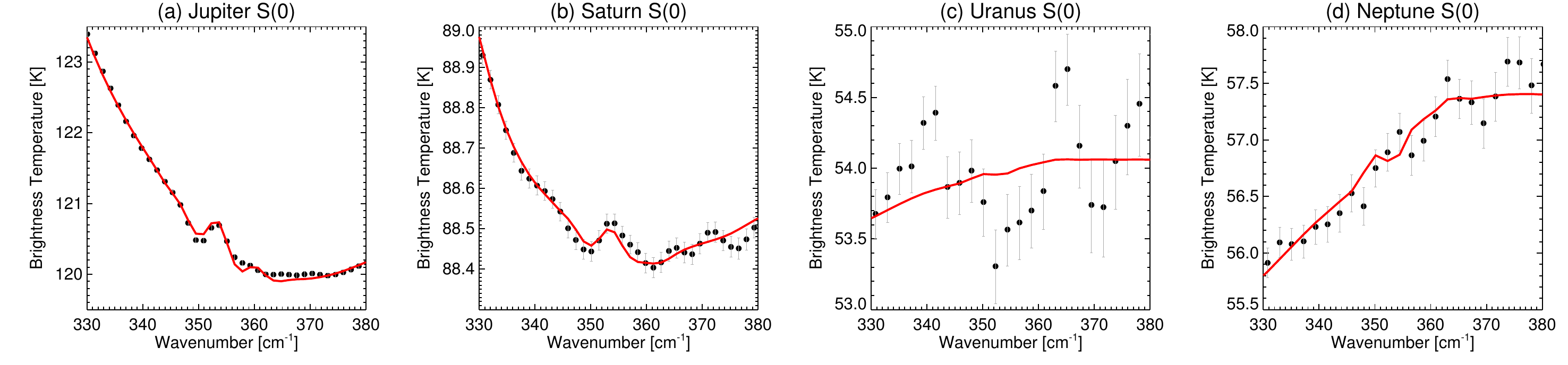}
\caption{Large averages of Voyager IRIS spectra (black circles with standard errors on the mean) compared to the spectral model with all dimer transitions included (red line).  Dimer absorption is clearly visible for Jupiter and Saturn observations, whereas the features are indistinguishable above the noise for Uranus and Neptune.}
\label{compiris}
\end{figure*}

The quality of the IRIS spectral fits using the latter database is shown for all four planets in Fig. \ref{compiris}.  For the purpose of this plot, we averaged all Voyager-1 IRIS Jupiter and Saturn spectra within $\pm30^\circ$ latitude of the equator, whereas for Uranus and Neptune we averaged all available Voyager-2 IRIS spectra from pole to pole, filtering for any corrupted measurements.  Dimer features near $S(0)$ are clearly identifiable and well reproduced for Jupiter and Saturn \citep{84frommhold}, but are indistinguishable above the IRIS noise for Uranus and Neptune.

\begin{figure*}[ht!]
\includegraphics[width=\linewidth]{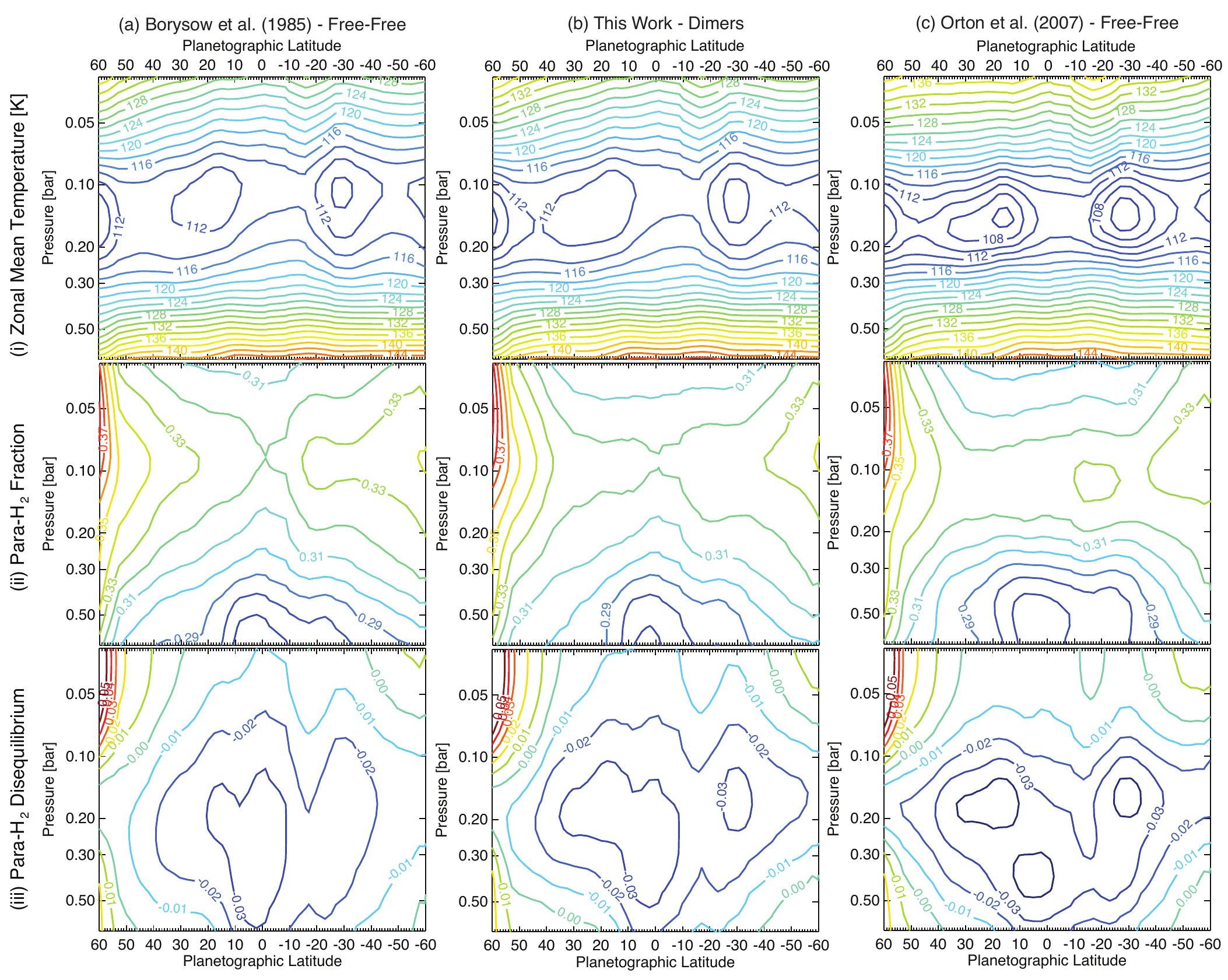}
\caption{Comparison of retrieved (i) tropospheric temperatures; (ii) para-H$_2$ fraction; and (iii) difference between para-H$_2$ and equilibrium.  More negative values in the bottom row indicate more para-H$_2$ than expected from equilibrium (i.e., sub-equilibrium conditions), potentially associated with upwelling motions.  The results are compared between calculations using (a) the free-free continuum absorption of \citet{85borysow}, (b) the new compilation with dimer absorption; and (c) the free-free continuum absorption of \citet{07orton}.}
\label{jupiris}
\end{figure*}

Fig. \ref{jupiris} shows the consequences of using the three different CIA databases on retrievals of temperature and para-H$_2$ ($f_p$) from Voyager IRIS Jupiter spectra.  As shown in \citet{17fletcher_sofia}, the use of the free-free continuum of \citet{07orton} (Fig. \ref{jupiris}c), without the additional absorption from the dimers near the S$(0)$ and S$(1)$ peaks, results in temperature retrievals with sharp lapse rates and low temperatures ($T<106$ K) near the tropopause.  In addition, the para-H$_2$ fraction is larger ($f_p=0.33-0.34$) as the spectral model attempted to increase the radiance at 350 cm$^{-1}$ by increasing the $f_p$, resulting in a strongly sub-equilibrium atmosphere (i.e., para-H$_2$ exceeds that expected from equilibrium).  The goodness-of-fit was also significantly worse in the free-free case, by a factor of three. Note that we detected a $0.9$-cm$^{-1}$ offset between our spectral model and the Voyager-1 IRIS Jupiter data \citep[which was also present in the analysis of][]{17fletcher_sofia} which was not present in the Cassini and Spitzer comparisons.  The IRIS data in Fig. \ref{compiris} have been shifted in wavenumber to compensate.  With the new dimer database, the retrieved tropospheric temperatures are smoother, with a minimum $T\sim110$ K and slightly lower $f_p=0.32-0.33$ leading to para-H$_2$ conditions closer to equilibrium \citep[albeit perturbed by equatorial upwelling and polar subsidence, as described by][]{98conrath, 17fletcher_sofia}.  This comes close to matching the calculations using the original database of \citet{85borysow}, where some of the free-free components had been overestimated.

The same exercise was repeated for the IRIS Saturn data \citep{16fletcher}.  Although the temperatures retrieved using the three compilations were similar (within $\sim1$ K), the para-H$_2$ fraction altered by $\sim0.03$ in the upper troposphere.  As was the case for Jupiter, this caused the calculation with only the free-free database of \citet{07orton} to appear strongly sub-equilibrium (i.e., more para-H$_2$ required to increase the absorption near S$(0))$), whereas the dimeric absorption fulfils this role in our new calculation, bringing the atmosphere closer to equilibrium, albeit with seasonal north-south gradients in Saturn's para-H$_2$ described by \citet{16fletcher}.  For Uranus and Neptune, where only the 200-400 cm$^{-1}$ offer any constraint (and there is no sensitivity to the S$(1)$ at all), the differences in the retrieved temperatures and para-H$_2$ were negligible and the dimer structure near 354 cm$^{-1}$ was not visible in the Voyager/IRIS 4.3-cm$^{-1}$-resolution data (Fig. \ref{compiris}c-d).




\section{Conclusions}
\label{conclusion}

The purely free-to-free H$_2$-H$_2$ collision-induced absorption tables of \citet{07orton}, which are included in the `Alternate' directory of HITRAN 2012 \citep{12richard}, should not be used in isolation when fitting mid- and far-infrared spectra in the vicinity of the S$(0)$ and S$(1)$ lines.  Instead, we propose that they should be combined with the present refined opacity offered by free-to-free, free-to-bound, bound-to-free and bound-to-bound H$_2$-H$_2$ transitions within $\pm50$ cm$^{-1}$ of the line centres. A new $\textit{ab initio}$ model, using the isotropic interaction potential approximation and a denser energy grid than previous studies, is used to calculate the opacity provided by the former three types of transitions for a range of temperatures and para-H$_2$ fractions.  Similarly, the opacity contribution from bound-to-bound transitions is computed with the full anisotropic interaction potential.  This model is validated via comparison to laboratory measurements at 20 and 77 K \citep{91mckellar,88mckellar}, and then used in radiative transfer calculations to simulate the spectra of the four giant planets.

We find that we can reproduce the low-resolution (4.3 cm$^{-1}$) dimer signatures that had been previously identified in Jupiter and Saturn Voyager spectra by \citet{84mckellar}, and that the additional continuum absorption resolves the problems that were identified by \citet{17fletcher_sofia} when using the free-free calculations of \citet{07orton} in isolation.  Observations at higher spectral resolutions (0.5-1.0 cm$^{-1}$) were compiled from Cassini Composite Infrared Spectrometer (CIRS) observations of Jupiter and Saturn and Spitzer Infrared Spectrometer (IRS) observations of Uranus and Neptune.  The model-data comparison reveals the presence of dimer absorption near S$(0)$ on all four planets at high spectral resolution, and near S$(1)$ on Uranus and Neptune (the structure is lost in the data uncertainty on Jupiter and Saturn).  Dimeric transitions appear in absorption on the gas giants, where they sense the upper troposphere, and in emission on the ice giants, where they sense the lower stratosphere.

Fits to the data are adequate but not perfect.  For example, the importance of bound-to-bound transitions is unclear, as they improve the fit for Jupiter but not for the other giant planets.  Secondly, both Uranus and Neptune have model-data discrepancies between the quadrupole at 354.5 cm$^{-1}$ and the primary dimer feature at 351 cm$^{-1}$ that we cannot account for using the \textit{ab initio} model.  It should be noted that the spectral calculations described in this work account only for binary interactions (between two hydrogen molecules), and not ternary or higher-order interactions.  The sharp dimer features correspond to quantum states that are rather long-lived, and thus may be sensitive to these additional interactions, which is a possible source of error in the present modelling work.  Currently, ternary and higher-order interactions cannot be treated well with quantum mechanics, which would be necessary in the case of hydrogen.  One has to resort to classical mechanical treatments \citep[see, e.g.,][]{11hartmann,17fakhardji}, which are typically well-suited for higher mass molecules such as CO$_2$ or N$_2$.  Nevertheless, the binary theory and isotropic potential approximation are able to reproduce much of the fine-scale structure observed in the giant planet spectra presented in this work.

It is possible that better observational data, with a higher sensitivity and spectral sampling, will help resolve these discrepancies.  The James Webb Space Telescope, when it launches in 2019, will provide exquisite spectral capabilities in the 5-30 $\mu$m range using the MIRI integral field units \citep{15wells}. Unfortunately, the collision-induced continuum from Jupiter and Saturn are likely to cause detector saturation (although wavelengths shortward of 11 and 16 $\mu$m are likely to be accessible on Jupiter and Saturn, respectively).  However, no saturation is expected for Uranus and Neptune, where both S$(0)$ and S$(1)$ will be available for further study at spectral resolutions of $R\sim3000$ (channel 3LONG, 15.4-18.1 $\mu$m) and $R\sim1600$ (channel 4LONG, 24.0-28.5 $\mu$m).  Observations of all four giant planets are scheduled as part of the Guaranteed Time Program of H. Hammel and will provide an excellent test of the new dimer spectra provided in this study.  Finally, given the importance of this spectral range in determining the helium abundances on the giant planets via remote sensing \citep[e.g.,][]{00conrath}, we hope that this new dimer spectral database will be of use to the wider community.



\acknowledgments

Fletcher was supported by a Royal Society Research Fellowship and European Research Council Consolidator Grant (under the European Union's Horizon 2020 research and innovation programme, grant agreement No 723890) at the University of Leicester.  Gustafsson acknowledges support from the Knut and Alice Wallenberg Foundation.  Orton was supported by funding from the National Aeronautics and Space Administration to the Jet Propulsion Laboratory, California Institute of Technology.    The new free-to-free, free-to-bound, bound-to-free and bound-to-bound opacity tables are all available at the following address:  \url{https://doi.org/10.5281/zenodo.1095503}.

%

\vspace{5mm}
\facilities{Cassini, Spitzer, Voyager}





\appendix
\section{Bound-to-bound spectra}
\label{appendix_bb}

The bound states are computed with a discrete variable representation (DVR) on a uniform grid \citep{92colbert}.  The formulas outlined below are for the case of two distinguishable hydrogen molecules, i.e., a complex consisting of one para-H$_2$ and one ortho-H$_2$.  At the end we will describe the modifications needed to treat a complex consisting of two identical hydrogen molecules.  We will also consider the hydrogen molecules as rigid rotors, as they are throughout this work.  Under those conditions the Hamiltonian for two hydrogen molecules, with centres of mass separated by the vector $\bf R$, is:
\begin{equation} \label{ehamiltonian}
H(\widehat{\bf r}_1, \widehat{\bf r}_2, {\bf R}) = H^{\rm mol}(\widehat{\bf r}_1)
   + H^{\rm mol}(\widehat{\bf r}_2)
   - \frac{\hbar^2}{2 m}\nabla^2_{\bf R}
   + V(\widehat{\bf r}_1, \widehat{\bf r}_2, {\bf R})
\end{equation}
where the orientations of the two diatoms are given by the unit vectors $\widehat{\bf r}_1$ and $\widehat{\bf r}_2$.  $V$ is the interaction potential and $m$ is the bimolecular reduced mass.  The wave function is expanded in angular basis functions according to:
\begin{eqnarray} \label{ebasis}
\Psi_{JM}(\widehat{\bf r}_1, \widehat{\bf r}_2, {\bf R})
= \sum_{\beta}
   \frac{1}{R} \; F^J_{\beta}(R) \,
   Y^{J M}_{j_1 j_2 j l}(\widehat{\bf r}_1,
   \widehat{\bf r}_2, \widehat{\bf R})
\end{eqnarray}
where $\beta$ is shorthand for the quantum numbers $(j_1 j_2 j l)$ and $Y^{J M}_{j_1 j_2 j l}$ is the vector coupling function:
\begin{eqnarray} \label{e3vect}
Y^{J M}_{j_1 j_2 j l}(\widehat{\bf r}_1, \widehat{\bf r}_2,
   \widehat{\bf R}) =
   \sum_{m_1 m_2 m_j m_l}
   C(j_1, j_2, j; m_1, m_2, m_j) \, C(j, l, J; m_j, m_l, M) \nonumber \\
   \times Y_{j_1 m_1}(\widehat{\bf r}_1) \,
   Y_{j_2 m_2}(\widehat{\bf r}_2) \,
   Y_{l m_l}(\widehat{\bf R}) \; ,
\end{eqnarray}
corresponding to ${\bf J} = {\bf j} +{\bf l}$ and ${\bf j} = {\bf j}_1 + {\bf j}_2$.  The angular momentum quantum numbers ${\bf j}_1$ and ${\bf j}_2$ correspond to the diatomic rotations, and ${\bf l}$ and ${\bf J}$ are the end-over-end and total angular momenta, respectively.  The factors $C$ are Clebsch-Gordan coefficients. The potential is also expanded in spherical harmonics as in \citet{03gustafsson} and then the expansion (\ref{ebasis}) and the Hamiltonian (\ref{ehamiltonian}) yields the DVR-matrix Hamiltonian:
\begin{eqnarray} \label{eHdvr}
  H^J_{\alpha\alpha'\beta\beta'} = \left(
  E^{\rm mol}_{j_1} + E^{\rm mol}_{j_2}
  + \frac{\hbar^2 l(l+1)}{2 m R_\alpha^2} \right) \delta_{\alpha\alpha'} \delta_{\beta\beta'}
  + T_{\alpha\alpha'} \delta_{\beta\beta'}
  + V^J_{\beta\beta'}(R_\alpha) \delta_{\alpha\alpha'}
\end{eqnarray}
with the kinetic energy matrix:
\begin{equation}
T_{\alpha\alpha'} = \frac{\hbar^2}{2 m \, \Delta R^2} (-1)^{\alpha-\alpha'}
\left\{ \begin{array}{ll}
\frac{\pi^2}{3} - \frac{1}{2 \alpha^2} \;& \alpha = \alpha' \\
\frac{2}{(\alpha - \alpha')^2} - \frac{2}{(\alpha + \alpha')^2}
 \;& \alpha \neq \alpha'
\end{array}\right.
\end{equation}
where the uniform DVR grid has a spacing $\Delta R = (R_{\rm max} - R_{\rm min})/N$ and corresponding grid points:
$R_\alpha = R_{\rm min} + \alpha \cdot \Delta R$
where $\alpha = 1,2...(N-1)$.
The Hamiltonian in Eq.~(\ref{eHdvr}) is independent of the quantum number $M$ \citep{75green} and thus the eigenvectors and eigenenergies are also $M$-independent. The H$_2$ rotational energies $E^{\rm mol}_{j_i}$ are taken from \citet{57stoicheff} and they are shifted so that $E^{\rm mol}_0 = 0$.  The potential matrix element is:
\begin{equation} \label{eVH2H2}
V^J_{\beta\beta'}(R_\alpha) =
\sum_{\gamma_1 \gamma_2 \gamma}
V_{\gamma_1 \gamma_2 \gamma}(R_\alpha) \;
e_{\gamma_1 \gamma_2 \gamma}(j_1, j_2, j, l, j_1', j_2', j', l'; J) \; ,
\end{equation}
where the coefficient, $e_{\gamma_1 \gamma_2 \gamma}$, and the expansion of $V$
is given in \citet{03gustafsson}.  The Schr\"odinger equation is solved through diagonalisation of the Hamiltonian (\ref{eHdvr}) and $(N-1)N_\beta$ eigenvectors, $F^J_{\beta k}(R_\alpha)$, and eigenenergies, $E^J_k$ are obtained.  $N_\beta$ is the number of angular momentum basis functions in the expansion, Eq.~(\ref{ebasis}).  The diagonalisation is done with the DSYEV routine from LAPACK \citep{99anderson}.

With the eigenstates determined as described above the absorption coefficient can be computed from the matrix elements of the interaction-induced electric dipole moment.  The temperature dependent absorption coefficient is (see e.g.~\citet{15karman}):
\begin{eqnarray} \label{eabs}
\frac{\alpha(\nu, T)}{\rho^2} &=& \sum_{E^{J'}_{k'} > E^J_k} \frac{4 \pi^3}{hc} \nu^{J'J}_{k'k} \left(1 - \exp\left(-\frac{E^{J'}_{k'} - E^J_k}{k_B T}\right)\right)
\frac{h^3}{(2 \pi m k_B T)^{3/2}} \, g^\epsilon_{j_1 j_2}(f_p) \nonumber \\ &\times& \exp\left(-\frac{E^J_k-E_{\rm asympt}}{k_B T}\right) 
\left|M^{JJ'}_{kk'}\right|^2 \left(\frac{1}{w} - \frac{\max(\nu - \nu^{J'J}_{k'k}, \nu^{J'J}_{k'k} - \nu)}{w^2} \right)
\end{eqnarray}
where $\nu$ is the wavenumber of the radiation in cm$^{-1}$ and $\nu^{J'J}_{k'k} = \frac{E^{J'}_{k'} - E^J_k}{hc}$ is the transition wavenumber. $E_{\rm asympt}$ is the lowest asymptotic energy when the two hydrogen molecules are separated, i.e.\ for the case of one ortho- and one para-hydrogen $\frac{E_{\rm asympt}}{hc} = \frac{E^{\rm mol}_1}{hc} \approx 118.5$ cm$^{-1}$. The last parenthesis in Eq.\ (\ref{eabs}) produces a triangular line profile of width $w$.  The fraction of para-H$_2$ in the gas is $f_p$, which is 1, 0.5, 0.25 for pure para-H$_2$, equilibrium-H$_2$ at 77~K, and normal-H$_2$, respectively.  Those are the cases considered in the comparison with laboratory measurements in section~\ref{calc}.  The statistical weight $g$ is given in table~\ref{tweights}.  The matrix elements of the dipole moment is:
\begin{equation} \label{edipmatrix}
M^{JJ'}_{kk'} = \sum_\beta \sum_{\beta'} \sum_\alpha \sum_{\lambda_1 \lambda_2 \lambda L}
F^J_{\beta k}(R_\alpha) \, A_{\lambda_1 \lambda_2 \lambda L}(R_\alpha) \, F^{J'}_{\beta' k'}(R_\alpha) \, d_{\lambda_1 \lambda_2 \lambda L}^{\beta J \beta' J'}
\end{equation}
where the coefficient $d_{\lambda_1 \lambda_2 \lambda L}^{\beta J \beta' J'}$ is given in \citet{03gustafsson}.  Note that the coefficient $d_{\lambda_1 \lambda_2 \lambda L}^{\beta J \beta' J'}$ includes a factor $1/\sqrt{3}$, which would otherwise have appeared as 1/3 in Eq.~(\ref{eabs}).  The spherical dipole components $A_{\lambda_1 \lambda_2 \lambda L}$ come from the expansion of the electric dipole moment:
\begin{equation}
\mu_z (\widehat{\bf r}_1, \widehat{\bf r}_2, {\bf R})
 = \frac{(4 \pi)^{3/2}}{\sqrt{3}} \;
   \sum_{\lambda_1 \lambda_2 \lambda L}
   A_{\lambda_1 \lambda_2 \lambda L}(R) \;
   Y^{10}_{\lambda_1 \lambda_2 \lambda L}(\widehat{\bf r}_1, \widehat{\bf r}_2,
   \widehat{\bf R})
\end{equation}
where only the $z$-component has to be considered because the $z$-direction defines the angular momentum quantization axis in a space fixed frame of reference \citep{82julienne}.

\begin{table} 
  \centering
  \caption{Statistical weights $g^\epsilon_{j_1 j_2}(f_p)$ for different combinations of para- and ortho-H$_2$ and different
  symmetries $\epsilon$} \label{tweights}
  \renewcommand
      {\tabcolsep}{8pt}
      \begin{tabular}{ccc|c}
        \hline\hline
        $j_1$ & $j_2$ & $\epsilon$ & $g^\epsilon_{j_1 j_2}(f_p)$  \\
        \hline
        even & even & $+$ & $f_p^2$ \\
        even & even & $-$ & 0 \\
        odd & odd & $+$ & $\frac{2}{3}(1 - f_p)^2$ \\
        odd & odd & $-$ & $\frac{1}{3}(1 - f_p)^2$ \\
        even/odd & odd/even &  & $f_p (1 - f_p)$ \\
        \hline\hline
      \end{tabular}
\end{table}

For the case of dimers consisting of two identical molecules (para-para or ortho-ortho) the wave functions have to be symmetrized.  The symmetry parameter $\epsilon$ is $+$ and $-$ for symmetric and antisymmetric wave functions, respectively. Two of the formulas above have to be modified accordingly. In Eq.~(\ref{eVH2H2}) the coefficient $e_{\gamma_1 \gamma_2 \gamma}$ has to be replaced by its $\epsilon$-dependent version, and the same goes for $d_{\lambda_1 \lambda_2 \lambda L}$ in Eq.~(\ref{edipmatrix}).  The $\epsilon$-dependent coefficients are given in the appendix of \citet{03gustafsson}.  Finally, the dipole selection rules have to be obeyed for all transitions included in the calculation of the absorption coefficient according to Eqs.~(\ref{eabs}) and (\ref{edipmatrix}).  These are the following:
\begin{itemize}
\item Total angular momentum:  $J' = J ; \; J \pm 1$.
\item $j_i$ and $j'_i$ are both even or both odd for $i$=1 or 2.
\item Parity must change, implying that $l$ changes from odd to even or vice versa.
\item In the case of identical molecules the symmetry, $\epsilon$, is conserved.
\end{itemize}




\bibliographystyle{aasjournal}
\bibliography{references.bib}

%
%



\end{document}